%% file: mainZac.tex
\documentclass[a4paper,10pt]{article}
\usepackage[utf8]{inputenc}
\usepackage{graphicx,authblk}
\usepackage[margin=1.5cm]{geometry}
\usepackage{color}
\definecolor{shadecolor}{rgb}{1,0.8,0.3}
\definecolor{TFFrameColor}{rgb}{1,0.8,0.3}
\definecolor{TFTitleColor}{rgb}{0,0,0}
\usepackage{xcolor}
\usepackage{subfigure}
\usepackage[round, sort, numbers, authoryear, sectionbib]{natbib}
\bibpunct{(}{)}{,}{}{,}{,}

\title{Complex economies have a lateral escape from the poverty trap} 
\author{Emanuele Pugliese, Guido L. Chiarotti, Andrea Zaccaria\footnote{and.zaccaria@gmail.com}, and Luciano 
Pietronero}
\affil{ISC-CNR, Via dei Taurini 19, 00185, Rome, Italy}
\affil{Dipartimento di Fisica, Sapienza Universit\`a di Roma, P.le Aldo Moro 2, 00185, Rome, Italy}

\begin{document}
\graphicspath{{figures/}} 

\maketitle

\begin{abstract}
\input{abstract}
\end{abstract}

\section{Introduction}

The industrialization of a country is an impressive process, deeply changing 
the population and the institutions of the country while new and old resources 
are tapped to achieve growth. 
During this transition the growth rate of the economy is much higher than 
the global average and much higher than the past and future growth rate of that
country. It is however a transition and, as such, limited in time. When the 
process of 
industrialization has touched all the sectors, when the population is educated 
and near full employment, when all the scale economies have been fulfilled, 
the 
process loses its revolutionary power and the new society now sits among the 
developed countries.

Two questions haunted economists since the beginning of the discipline, since 
Adam 
Smith and Max Weber. The first question is about the drivers of this sudden 
sprout of growth,
how an entire society changes dramatically in fifty years after thousands of 
years 
of stillness. Related to this, why the process is as suddenly interrupted while 
the growth slows down after catching-up with the other developed countries.
There are many competing answers to this part of the puzzle, the most basic 
being the poverty trap due to multiple equilibria that was already in 
\cite{solow1956contribution}: the phenomenon is sudden because there is a 
barrier, a country wealth that has to be reached to start a quick transition to 
a different equilibrium.
Many alternative explanations are there, most notable increasing returns and 
demand, as in \cite{rosenstein1943problems} and - more formally - 
\cite{murphy1988industrialization}, in which the barrier to overcome is a 
minimum internal demand to allow for returns to scale in manufacturing.

The second question is about the heterogeneity of the process among countries. 
England experienced its Industrial Revolution
in the second half of the XVIII century, followed by other Western countries.
In the XIX century the United States industrialized 
too, and other countries followed in the XX century. Other countries did not.
What are the countries lagging behind missing to move toward prosperity? 
There are many alternative explanations, from cultural 
(\cite{weber2002protestant}, \cite{mccloskey2010bourgeois})
to geographic (\cite{diamond1997guns}), even biological (\cite{ashraf2011out}).
Another explanation is political : 
to achieve growth the population has to be empowered through inclusive 
political 
institutions, leading to more inclusive economic institutions and diffused 
prosperity (\cite{acemoglu2005institutions}). 

In this paper, we address both of the questions above by exploring the nature 
of the barrier to industrialization and its correlation to the complexity of the 
economy at the onset of the industrialization process. Indeed, along the 
industrialization path, the industrial capabilities of the country, the 
corresponding products and the consumer preferences, are completely reorganized, 
leading to the population’s freedom to pursue their own interests in unexpected 
(new) sectors and entrepreneurial activities. This fact dramatically increases 
the diversification and the complexity of the underlying economy.
We quantitatively measure the complexity of an economy through a new dimension, 
the fitness of the country, that has recently been introduced in the study of 
social and economic systems (Tacchella etal.2012). The latter share with 
traditional complex systems the emergence of unexpected collective behaviors 
coming from the non-trivial interactions between their basic components 
(Anderson 1972). The industrialization of a country is a dynamic process in 
which a complex network reinforcing production capabilities and product demand 
emerges at the country scale. The prosperity and the potential of a country can 
then be characterized (Cristelli et al. 2013) by considering this new dimension, 
which takes into account the diversification and the complexity of the 
production system.
In this paper, we investigate the role played by this new measure in the 
countries’ industrialization process. In order to perform this analysis we look 
at the empirical growth patterns of countries having different levels of 
fitness. As it will emerge, fitness, which is a quantity tuned on the 
diversification and complexity of the country’s export basket, carries important 
information with regard to the onset of the industrialization process. In 
particular we will see how a higher value of the fitness is associated with a 
lower barrier toward industrialization of a country. This empirical finding will 
motivate an attempt to build a full-fledged model of industrialization taking 
fitness endogenously into account as a proxy for the growth potential of 
countries. However, this will only be outlined in the conclusions as its 
mathematical formulation goes beyond the purpose of this paper.

%In the past years an approach based on methodologies from the Science of 
%Complex Systems have been used with 
%some success to characterize Social and Economic systems. 
%Indeed these systems share with the traditional objects of Complex Systems 
%analysis the emergence of an unexpected collective behavior coming from the 
%non 
%trivial interactions between components \citep{Anderson04081972}.
%The industrialization of a country is a dynamic process in which a complex 
%network reinforcing production capabilities and product demand emerges at the 
%country scale. 
%Complexity is then expected to play a crucial role in the emergence of growth 
%patterns. It has recently been suggested \citep{plosnm,newmetrics} that the 
%wellness and the potential of countries can be better characterized by 
%considering a further dimension, the 
%\emph{Fitness} of that country, which takes into account the diversification 
%and the complexity of the production system. 
%In this paper we investigate the role played by this new measure in describing 
%the countries’ industrialization process. 
%As it will turn out, fitness, which is a quantity tuned on the diversification 
%and complexity of the country's products export, carries an important 
%information with regard to the start of the industrialization process. 
%In particular we will see how an higher Fitness is associated with a lower 
%barrier to industrialize.

The following essay will be divided in four sections in addition to this 
introduction.
In the next section we will describe the 
economic 
complexity approach.
We will then briefly sketch a simple poverty trap model, a basic rendition of 
\cite{solow1956contribution}, to fix notation.
In the following section we will show some empirical shortcoming of this basic 
idea of poverty trap and we will show the promising role of our Fitness index 
to solve these shortcoming.
Finally, in section \ref{sec:model} we will show that introducing the Fitness 
in the basic poverty trap narrative helps the models to describe the empirical 
evidence.
The data used for our analysis is described in the appendix.

\section{The Neoclassical Economic approach to Growth}

\subsection{The poverty trap in standard economic theory \label{sec:Growth}}

In most of countries that will eventually join the other developed 
nations, the early stages of economic growth are characterized by a period of 
fast growth. The growth of the Soviet Union that scared the United States 
administration and economists \cite{Krugman1994}, the growth of Japan and 
Southern Europe in the ’50s, the growth of the Asian Tigers in the ’70s: every 
country emerging from of an agriculture based economy has experienced a decade 
or more of extremely high growth while it is catching up with the other 
developed countries. 
The spike of high growth is characterized by a strong increase in investments, 
both in physical and human capital. While the population experiences new 
incentives and opportunities for education and investment, the factors of 
production rise inflating the economic growth. This is what we define as 
industrialization: the moment in which there is a sudden spike in the factors of 
production available in the country. 

Since the influential \cite{solow1956contribution} this sudden transition from 
mere subsistence to a complete industrialization has been described with the 
presence of a trap that could be avoided by overcoming a barrier in terms of 
wealth, or physical capital. More formally the usual picture makes use of the presence of 
multiple equilibria in the time evolution of physical capital. In this picture, 
the presence of multiple equilibria is due to the non linear relation  between 
the country investments in physical capital and GDP  due to 
the presence of a threshold in the saving rate. 
The dynamic evolution results in a non linear increase of the available capital 
at successive times. The main ideas behind \cite{solow1956contribution} 
are presented in appendix \ref{apx:solow} for the interested reader.

This kind of explanation, being related to investments, is said ``supply side'', since it implies
that the the lack of enough investments is the
missing variable for achieving industrialization.  In this picture what is missing to inject growth are plants
able to produce more profits to be reinvested again: overcoming a threshold level of physical capital is necessary to 
start a expansive cycle.
There are many competing explanations however, typically marked as ``demand side'', 
where the lacking variable is enough internal demand to kick-start industrialization, 
\cite{rosenstein1943problems, murphy1988industrialization}.
In this kind of explanations, often referred as ``The Big Push'', one sector industrializing
increases the wages of part of the population, that requires more goods and, therefore, pushes 
for the industrialization of other sectors in a cascade dynamics.
While to address the issue of these different competing narratives is an important research 
question, it is not the question we want to answer in this analysis.

Therefore any explanation, both demand and supply side, can be assumed in the 
following, as long as it allows for a threshold of resources needed to start 
industrialization.
Since GDP per capita - required as a threshold by a demand side model of poverty 
trap with increasing returns - and the physical capita - required as a threshold 
by a traditional model of poverty trap - correlate very strictly\footnote{In 
our sample, over the period 1963-2000, the Spearman's rank correlation between 
GDP and physical capital is 96.2\%.}, we will be able to keep the same 
fuzziness also in the next, more empirical, analysis.
Economic theory is used in our empirical analysis only to 
separate the  contribution to growth due to innovation from that due to inputs, as illustrated 
in the next section. 
 
% This line of reasoning does not 
% explain why some nations start the industrialization process while others not, or
% why every country that industrialized followed a very different pattern, starting
% industrialization at very heterogeneous level of physical capital, at odds with 
% the idea of a threshold.
% Moreover this is an intrinsically unidimensional picture, since a country may 
% escape the poverty trap only by leveraging on invested capital. That actually 
% means that one may need a capital increase to produce an increase in capital to 
% reach the new equilibrium.
% 
% In this work we aim at introducing in this picture a new dimension, namely the 
% complexity of the economy of a country. This will allow the discernment of the 
% heterogeneous patterns of countries' industrialization.   

\subsection{Separating the components of GDP growth}

To understand the role played by the complexity of the economy in the 
industrialization process we decompose the relative growth of output of the 
country production structure (GDP) into its main components: the one related to 
variation of technological efficiency and those related to variation of inputs, 
namely asset capital and human capital. This decomposition is standard in 
economic literature since the seminal work of \cite{solow1957technical}, and
its mathematical formulation can be derived from the model presented in appendix 
\ref{apx:solow}.   
%
%\subsection{Decomposing Economic Growth}
%
%The simple functional shape in equation \ref{eq:CDGrowth} is useful to 
%empirically 
%quantify the different kinds of growth. Let us first define the labor input 
%$L$ as 
%the product of the number of employees ($E$) and their average human capital 
%($H$).
In this setting 
%the GDP per capita is equal to
%\begin{equation}
%\left(\frac{Y_{i,t}}{P_{i,t}}\right)_{i,t}=A_{i,t}\left(\frac{K_{i,t}}{P_{i,t}}
%\right)^\alpha\left(\frac{E_{i,t}}{P_{i,t}}H_{i,t}\right)^{1-\alpha},
%\end{equation}
%and therefore, defining with the lowercase letters the growth rates of the 
%respective 
the GDP per capita growth rate can be written as:
% uppercase variables and with the hat the division by population,
\begin{equation}
 {y}_{c,t}=a_{c,t}+\alpha{k}_{c,t} + (1-\alpha){e}_{c,t} + 
(1-\alpha) h_{c,t}. \label{eq:GrowthDec}
\end{equation}
where $a_{c,t}$ is the growth rate of GDP per capita due to the (exogenous) 
technological efficiency of the country $c$ at time $t$, $\alpha {k}$ the growth 
due to the increase of physical capital per capita, $(1-\alpha){e}$ the growth due to 
an increase of 
the labor force 
share in population, $(1-\alpha)h$ is the growth due to an increase the human 
capital (education) of workers of the country, and $\alpha$ is the output 
elasticity of capital, where the output elasticity of labor is assumed to be  
$(1-\alpha)$.
%Formula \ref{eq:GrowthDec} divide the growth rate of GDP per capita, 
%$\hat{y}$, 
%in its component:
%the exogenous growth rate, $a$, the growth rate of GDP due to per capita 
%physical capital 
%accumulation, $\alpha \hat{k}$, the growth rate of GDP due to an increase of 
%the labor force 
%share in population, $(1-\alpha)\hat{e}$, and the growth rate of GDP due to an 
%increase 
%of the average human capital of workers, $(1-\alpha)h$.
These last three addenda form our definition of input growth, being it a 
physical investment
in new machinery ($k$), an increase in labor force participation($e$), or 
additional education ($h$).

Since the growth rate of inputs is quantifiable, to compute the different parts 
of growth in equation \ref{eq:GrowthDec} 
we need only to estimate $\alpha$.
Economic theory is handy in this case. If each factor of production is paid for 
at its marginal value, the
share of national income going to capital will be $\alpha$ and the share going 
to labor $1-\alpha$.
Since these shares are observable numbers, we will use them to estimate 
$\alpha$. 
Finally, the efficiency part $a$  can be recovered as 
the residual after removing the inputs component from the total GDP per capita 
growth.

While the overall measure of GDP growth can be deceiving and influenced by the 
price and discovery of natural resources and the happening of any external factor, 
the sudden investments in physical and human capital and the increase in labor force 
participation are the clear fingerprints of a structural change, a movement from an 
equilibrium to another.

\subsection{The growth due to inputs}

Why are we interested in input growth, and not in total per capita GDP growth? 
Much of the academic world focused only on efficiency and productivity, 
assessing that most of the long term growth is due to productivity growth (Hall 
and Jones 1999). This is obvious: input growth is intrinsically limited. While a 
country can double its employment rate from 30\% to 60\% in the first 20 years 
of industrialization, it cannot double it again in the following 20 years. While 
a country can quickly increase literacy rate to 90\%, further efforts cannot 
give similar payoffs. 
Even if physical capital could, in principle, grow without bounds, its 
effectiveness in terms of increasing labor productivity would decline once 
technological progress is included in the description. As a consequence, input 
growth suffers from decreasing returns, and it cannot be the focus for long-term 
growth in developed countries. 
However, the growth due to inputs is indeed a powerful force at the onset of a 
country’s industrialization.

Analyzing for example the case of industrialization of Singapore during its 
high growth phase, the excess growth required to catch up with the developed 
world was obtained through input growth. In the year of maximum growth for 
Singapore, 1970, out of an impressive 11\% growth rate of real per capita GDP, 
8\% was due to input growth.

Even without decomposing the growth, the transforming effect of 
industrialization on Singapore is visible by simply looking at the changes in 
the descriptive statistics in the time span of one generation. In 1966, at the 
beginning of Singapore’s industrialization, 27\% of the country population was 
employed, and 39\% of the new entrants in the job market had no formal 
education. Only 16\% of the new entrants in the job market had at least a 
secondary degree. Furthermore, investments in physical capital were modest, with 
only 10\% of saving rates. The generation entering the job market in 1990 were 
confronted with a deeply changed country. Female workers had entered massively 
in the job market, and 51\% of the population was now working, almost twice as 
much as in the previous generation. Among entrants in the job market, only 10\% 
do not have any formal education (one fourth compared with the previous 
generation) and 54\% have at least a secondary degree (a threefold increase). 
Saving and investing has become common among the population, the savings rate 
has increased fourfold reaching 39\% (Krugman 1994, Mukhopadhaya 2002). Not 
surprisingly, in the same years Singapore’s GDP has increased by almost 9 times 
and the per capita GDP has increased by almost 6 times. Of this exceptional 
growth, only a modest 25\% (around 1\% per year, in line with developed 
countries), can be assigned to a growth in productivity (data from Penn Table 
8.1). In conclusion, the input growth defines the trend of per capita GDP 
growth, while the residual productivity increase acts mostly as noise over the 
growth signal.

\section{An Economic Complexity approach to Growth}

As mentioned in the introduction, we expect economies showing different levels 
of complexity (different fitness values) to behave differently in the 
industrialization process. The idea that economic indicators can be used to 
summarize the affluence and growth potential of nations dates back to the 
seminal work of \cite{kuznets1946national}, which introduced the GDP as a 
measure of nations’ productivity in the Thirties in order to better understand 
how to tackle the Great Depression. Since then a notable number of economic 
indicators have been proposed (\cite{stock1989new}). In recent years particular 
attention has been dedicated to non-monetary indicators 
(\cite{costanza2009beyond,diener1997measuring}).  
All these indicators look deeply in the characteristics of the 
economy to unveil the hidden potential of a single country. However in a 
complex system, as for the economic system of a country, the characteristics of 
single elements are usually less important than their mutual interactions. A 
stark example is the evaluation of the relative authority of webpages done by 
Google through the PageRank algorithm (\cite{Brin_Page}), in which the ranking 
reflects the proprieties of the reference network connectivity. Modern goods 
markets constitute a network of products similar to the one formed by the nodes 
of the world wide web: ranking algorithms can then be efficiently used to 
characterize the network properties, and in particular to rank nations 
according to their manufacturing capabilities and product complexity. In other 
words, the basket of manufacturing capabilities of each country, which is 
representative of the social, cultural and technological structure of the 
underlying economy (\cite{lall1992technological,dosi2000nature}) 
(intangible and not directly measurable) is expressed by 
the basket of products the country is able to produce 
(\cite{hidalgo2007product}). 

\cite{hidalgo2009building} have pioneered this kind of approach for the 
economic science, and kindled a rich literature on the subject. In particular 
\cite{newmetrics} applied this view to introduce a novel non-monetary 
indicator based on the properties of the network formed by interstate goods 
exchanges in which non-linear interactions play a fundamental role. In this 
work we will follow this later approach. The information of the country growth 
potential can be extracted from the properties of the worldwide export network 
as expressed by the structure of the matrix \textbf{M} whose entries $M_{cp}$ 
take the value 
1 if the country $c$ shows a Revealed Comparative Advantage 
(\cite{balassa1965trade}) in the export of a (physical) product $p$ and 0 
otherwise. This choice has a number of advantages: i) any trivial correlation 
with export volumes is removed, thus allowing comparison of countries of 
different size; ii) the method allows comparison of countries with very 
different shares of manufacturing sectors of total GDP, since only its 
composition is relevant, while the aggregate volumes of export does not matter. 
Finally iii) while the matrix “binarization” 
necessarily implies a loss of information, it has been shown that this kind of 
treatment is far more resilient with respect of the possible presence of noise 
in the data (\cite{battiston2014metrics}). To measure the country export 
competitiveness (directly related to the country’s capabilities, in the spirit 
of \cite{hausmann2007you}) using only the above defined export network linking 
countries and products, could be sufficient to take into account the number of 
products exported by a given country, i.e. its diversification. However this 
would be simplistic, since one cannot assume that products are all equals in 
terms of the capabilities required to produce them: some products may in fact 
require a higher number of (or more exclusive) capabilities to be produced (and 
exported). In this sense we define a product to be complex if it requires a 
higher number of capabilities (or a more sophisticated set of capabilities) to 
be produced. Indeed what is required is a way to measure, although indirectly, 
the countries’ capabilities signaled by the production (and export) of a given 
product. Being the capabilities not directly measurable, we will summarize the 
capabilities needed to export a single product as a single value, that we call 
product complexity, computed only relying on the structure of the 
country-product network. Let us suppose for the moment this numeric value to be 
known. Given these product complexities, the fitness of a country will simply 
be defined as the sum of the complexities of its exported products 
(mathematical expression below) and it will be representative of the country’s 
capabilities. We will show in the following how this measure of fitness 
empirically relates to the industrialization of the country. In turn, product 
complexity can be defined through the fitness of the countries exporting them: 
a product is considered complex if low fitness countries do not export it. 
Consider this simple case: only highly developed countries export transistors, 
while both more industrialized and less industrialized countries export nails. 
Therefore the low complexity of nails can be deduced directly by the fact that 
low fitness countries are also able to produce and export them. As a 
consequence low fitness countries are more informative in assessing the 
complexity of products. All these considerations yield to a non-linear relation 
between fitness of countries and complexity of products.

More formally, to calculate the fitness of 
countries $F_c$ and the complexity of the exported products $Q_p$ we iterate 
upon convergence the following set of non-linear coupled equations: 

\begin{minipage}{0.45\textwidth}
\begin{equation}
 \tilde{F}_c^{(n)}=\sum_p M_{cp} Q_{p}^{(n-1)}
\end{equation}
\begin{equation}
 \tilde{Q}_p^{(n)}=\frac{1}{\sum_c M_{cp} \frac{1}{F_{c}^{(n-1)}}}
\end{equation}
\end{minipage}
\begin{minipage}{0.45\textwidth}
\begin{equation}
\label{jsdlskjfs}
 F_c^{(n)}=\frac{ \tilde{F}_c^{(n)}}{ <\tilde{F}_c^{(n)}>_c}
\end{equation}
\begin{equation}
\label{ljlkjljlkj}
 Q_p^{(n)}=\frac{ \tilde{Q}_p^{(n)}}{ <\tilde{Q}_p^{(n)}>_p}
\end{equation}
\end{minipage}

where the normalization of the intermediate tilted variables is made as a 
second step, n is the iteration index, and the $< >_x$ symbol stands for the 
arithmetic mean with respect to the possible values of $x$.

The fixed point of these maps has been studied with extensive numerical 
simulations and it is found to be stable and not depending on the initial 
conditions. We refer to \cite{plosnm} for a detailed description of 
the algorithm and a comparison with the one proposed by 
\cite{hidalgo2009building}. The convergence properties of the fitness and 
complexity algorithm are not trivial and have been studied in 
\cite{pugliese2014convergence}. This methodology has been applied to both the 
study of specific geographical areas, such as the Netherlands 
(\cite{zaccaria2015case}) and the Sub-Saharan countries 
(\cite{cristelli2015growth}), and general features of growth and development 
(\cite{cristelli2015heterogeneous, zaccaria2014taxonomy}).

\begin{figure}
\centering
 \includegraphics[width=15cm]{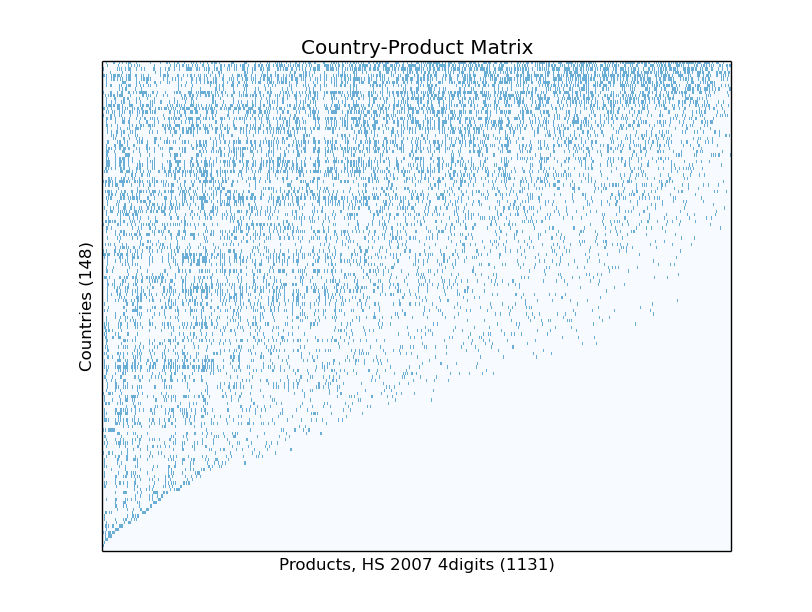}
 \caption{Binary matrix identifying countries producing a specific product. 
Countries (rows)
 are ordered according to their Fitness, Products (columns) are ordered 
according to their
 Complexity. A clear triangular structure emerges.}
 \label{fig:triangular}
\end{figure}

Once countries and products are arranged according to their respective fitness 
and complexity, the matrix $\textbf{M}$ is roughly triangular, as shown in 
\ref{fig:triangular}. This structure implies that developed countries tend to 
have a highly diversified export basket, since they export both complex and not 
complex products, while poor or less developed countries export fewer and less 
complex products. While diversification may lead to an immediate (zero-order) 
estimate of the fitness of a country commensurate with the number of different 
products it exports, the evaluation of the products’ complexity is more subtle: 
a product exported by low-fitness countries should be assigned a lower score 
since it is reasonable to expect that a lower level of capabilities is required 
to produce it. Clearly a linear page-ranking type of analysis cannot handle 
this point, while a nonlinear ranking approach – as the one presented above – 
will be more appropriate. 

\section{Empirical findings and their relation to poverty trap}

While economic theory is mostly related to equilibrium processes, studying 
balanced growth at the equilibrium ratio, industrialization is obviously a 
dynamic process occurring between different growth paths, while the country 
moves from one equilibrium ratio to another. Although growth in equilibrium is 
driven by the growth in productivity, the idea that there can be multiple 
equilibrium ratio of balanced growth and a barrier to overcome to shift from one 
equilibrium ratio to another was already present in Solow (1956). In the 
transition the country experiences high GDP growth through input growth. There 
can be a capital barrier, like in Solow (1956), or a demand barrier, like in 
Murphy et al. (1988). Still, there is a threshold to overcome in order to access 
the input driven out-of-equilibrium growth spike. Even in models considering the 
evolution of a country as a unified process, like Galor and Weil (2000), there 
are variables that must reach a tipping point in order to move the society into 
a high growth regime, driven by incentives to invest in production inputs. Given 
this, two stylized facts should then be expected to emerge by looking at 
empirical data: 

%What should we expect, looking at empirical data, from a poverty trap as 
%described in section \ref{sec:Growth}? 
%Two phenomena are expected.

First, if the catching up of 
developing countries is the result of the dynamics of inputs to a new 
equilibrium,
we should expect high input growth among the developing countries, sharply 
declining 
for the developed ones. We should therefore expect a negative relation between 
the GDP growth due to
input and the level of GDP for the countries that have started the transition: 
the 
growth should slow down for developing countries while the level of inputs 
approaches the new equilibrium and the developing countries catch up with the 
developed 
ones.

\begin{figure}[!ht]
 \includegraphics[width=\textwidth]{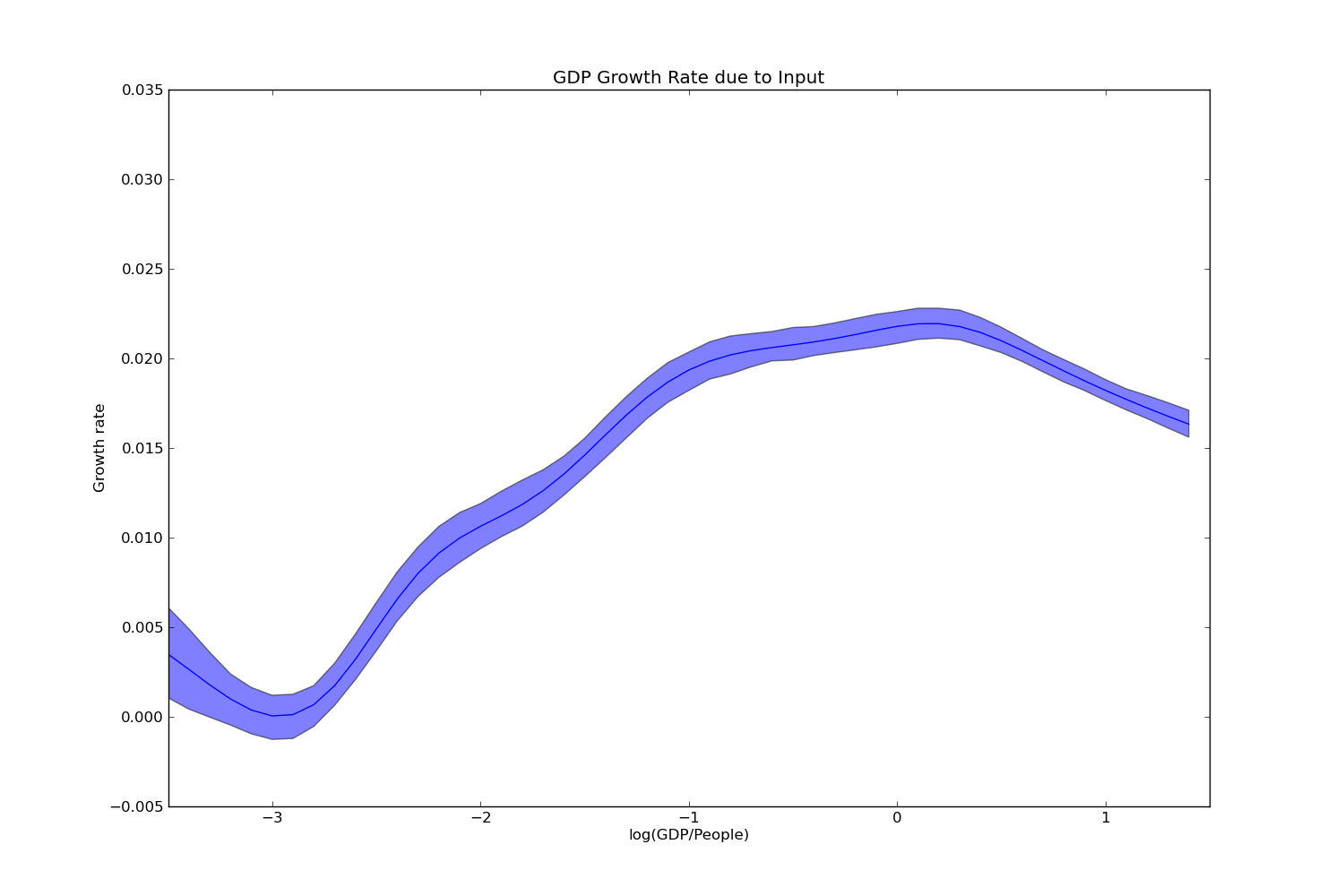}
 \caption{Non parametric kernel estimation of growth rate of per capita GDP due 
to inputs versus relative per capita GDP. The shadowing indicates 90\% of 
confidence interval of the expected value, computed with bootstrap. Different 
countries-years in the range 1963-2000 have been pooled after removing the 
global trend. While the low performance of low GDP countries in increasing their 
input is clearly visible (left site of the figure), the slowing down of input 
growth expected after catching-up (right side) is modest}
 \label{fig:GDPgrowthvsGDP}
\end{figure}

Second, we should expect a certain level of GDP (or physical capital, in Solow's perspective) per 
capita to be required to trigger the transition if the barrier to start the 
industrialization is demand driven as in \cite{rosenstein1943problems} (or capital driven, in Solow's perspective \cite{solow1956contribution}). We should therefore find a 
positive relation between per capita GDP growth due to inputs and the level of 
per capita GDP for low levels of per capita GDP, where additional per capita GDP 
means additional internal demand and implies higher per capita physical capital. 
To check these expected empirical behaviors we compute the average 
growth rate due to input versus the country related per capita GDP. We do so by 
pooling all the countries and years. In particular, we will use a non parametric 
Gaussian kernel estimation (Nadaraya 1964, Watson 1964) to compute the expected 
value of the per capita GDP growth rate due to input and the corresponding 
confidence interval for different values of per capita GDP. We plot the results 
in figure \ref{fig:GDPgrowthvsGDP}. 

%
%Second, both if the barrier to start the industrialization is demand driven as 
%in \cite{rosenstein1943problems}
%or if it is capital driven as in \cite{solow1956contribution}, 
%we should expect that a certain level of GDP/physical capital per capita is 
%needed to start the transition. 
%We should therefore find a positive relation between per capita GDP growth due 
%to input and the 
%level of per capita GDP for low levels of GDP per capita, where additional GDP 
%per capita
%means additional internal demand.
%
%Looking back at figure \ref{fig:multipleequilibria}, we expect therefore to 
%observe a decline while the system reach $K_2^*$ of an input growth spike that 
%started around $K_3^*$.

The results in figure \ref{fig:GDPgrowthvsGDP} do support the existence of a 
barrier to growth, since 
there is for 
sure a certain role played by prosperity to kick-start investments.
Data do not seem however to support the first hypothesis:
if any catching up mechanism is visible from the data, it is not an 
awe-inspiring event. 
The slow down for very high level of GDP per capita, while statistically 
significant, 
has poor economic meaning due to the presence of the plateau at medium large 
levels. For sure it does not support the image of 
calm after
the storm that we tried to evoke in the previous sections.

Clearly, at this level our analysis is missing a crucial ingredient: we are not 
able to pinpoint the possibly different growth potentials among countries. 
%The 
%previous simple exercise was narrating a very simple story in which every 
%country is following the same Growth trajectory: it was poor in equilibrium 
%$K^*_1$, it passed through an homogeneous threshold $K_3^*$ to get finally to 
%the same end point $K^*_2$.

As it is well known, some countries have started an impressive growth 
process, from an industrial and a social point of view, while others simply 
rely on the exploitation of natural resources. 
For the same level of physical capital or GDP, two different countries could 
live a moment of intense investment and shared opportunities for the whole of 
the population, favoring investments both in physical and human capital, or a 
moment of stillness and complacence, often characterized by exploitive economic 
institutions and high inequality. 
We need a quantitative measure in order to discriminate among these and others 
situations; from a practical point of view, a new dimension to disentangle 
different economies, possibly independent from the ones which are usually taken 
into account in mainstream economics. 
We believe that concepts taken from the economic complexity approach may be of 
help. In particular we believe that the fitness of a country, being both a 
quantitative measure of the number and of 
the quality of capabilities of a nation and a measure of diversification in 
advanced and complex products, is a useful indicator of the potential for 
growth of the country.

\begin{figure}[!ht]
 \includegraphics[width=\textwidth]{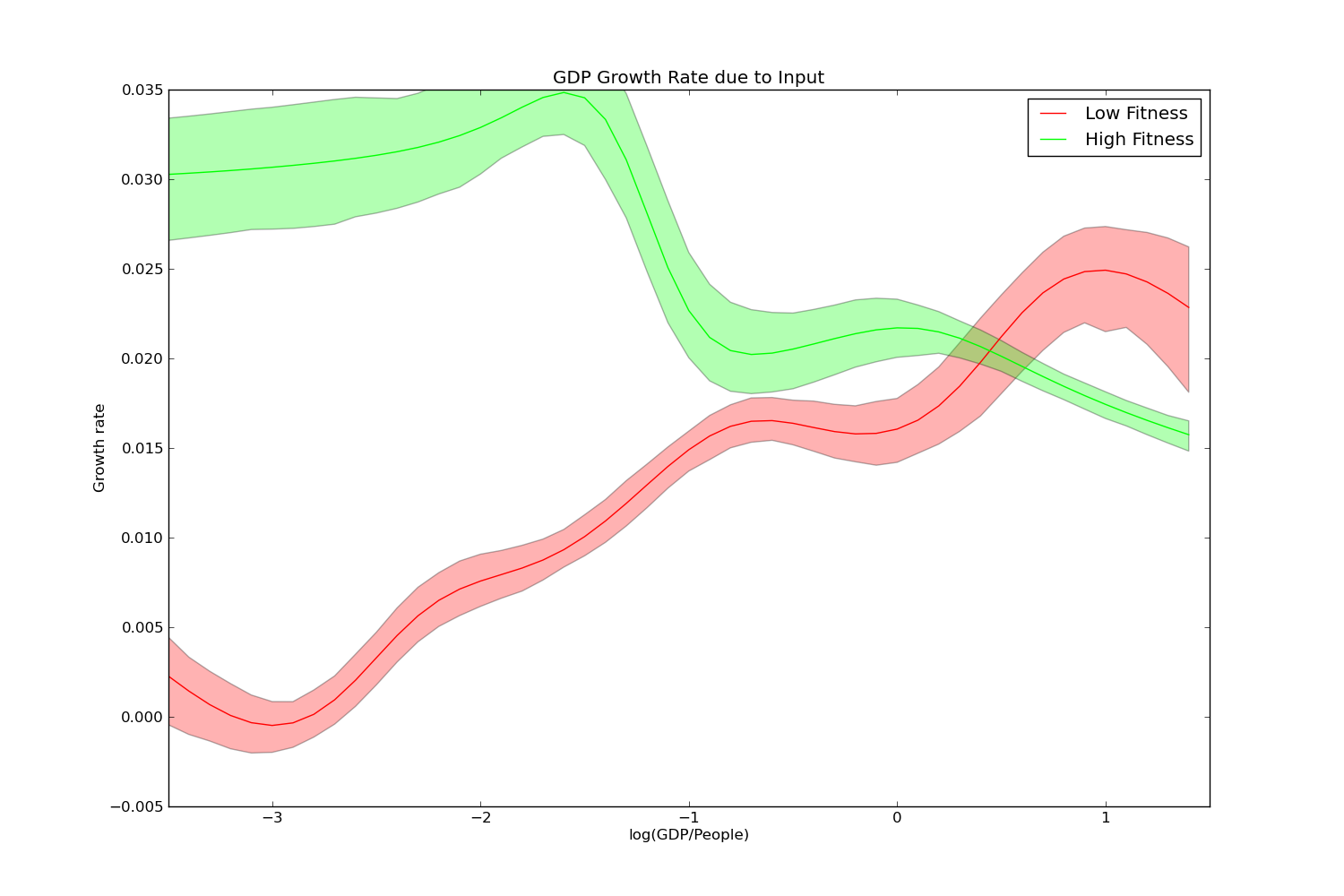}
  \caption{Non parametric Gaussian kernel estimation of growth rate of per 
capita GDP due to input versus per capita GDP for the lowest and the top tertile 
of the fitness distribution. The shadowing indicate the 90\% confidence 
interval of the expected value, computed with bootstrap. Different 
countries-years in the range 1963-2000 have been pooled after removing the 
global trend. Dividing the countries in sets depending on their fitness values 
highlights very different behaviors and reconciles the theory with the empirical 
observation.}
 \label{fig:GDPgrowthvsGDP_Bin}
\end{figure}

We will therefore replicate the exercise dividing the countries in three sets 
accordingly
to their fitness, to see if it helps disentangling the different regimes. 
In figure \ref{fig:GDPgrowthvsGDP_Bin} we show the results 
for the high fitness countries compared with the low fitness ones.

When data are split in this way, two different patterns emerge. What was 
confused when clubbing all
the countries together is now visible and in agreement with the predictions 
from section
\ref{sec:Growth}. The high fitness economies, the ones able to differentiate 
their production
in advanced products, present a clear downward slope of the growth of GDP per 
capita due 
to input growth with respect to the level of GDP per capita.
The countries with lower fitness instead have issues to start the transition. 
They
experience an higher barrier and they need a very high level of GDP per capita 
to experience
the mobilization of resources expected by economic theory.

The complexity based measure of country fitness seems to lower the barrier to 
start the transition, 
and we will see it in more detail in the next session.

\section{The role of complexity in Economic Growth}\label{sec:model}

\begin{figure}[!ht]
 \includegraphics[width=\textwidth]{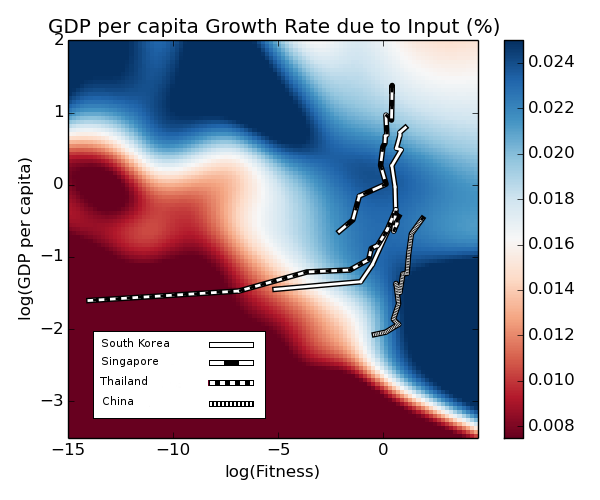}
 \caption{The color map represents the part Per Capita GDP Growth 
 due to inputs, for different values of Fitness and GDP per Capita.
 Different countries-years 
  in the range 1963-2000 have been pooled after removing the global trend.
  Both the role of the fitness of the country in lowering the threshold to 
enter in 
  the high endogenous GDP growth regime (the blue band in the center) and the 
slowing 
  down of the process for developed countries (the top-right corner) are 
evident.}
 \label{fig:IGvsGDP_3D}
\end{figure}

In the previous section we observed different behaviors for countries with 
different fitness levels, hinting at a possible explanation for the 
industrialization process of a country, which is fostered by high fitness 
levels. The complexity of the country’s economy brings down the barrier to 
industrialization and allows for investments in inputs. This hypothesis, 
however, requires further investigation. The different behavior of countries 
grouped according to their high (solid line) or low (dashed line) fitness levels 
(figure \ref{fig:GDPgrowthvsGDP_Bin}), needs to be generalized to a continuous description. 
This is the aim of the present section in which we compare the growth rate of 
per capita GDP due to inputs, with both the detrended  per capita GDP Y and 
the fitness F. This is achieved by a non parametric estimation of a two 
dimensional Gaussian kernel (Nadaraya 1964, Watson 1964) obtained by pooling all 
the countries and years for the time period in question. At difference with the 
analysis in figure \ref{fig:GDPgrowthvsGDP_Bin}, in which we compared only the behavior of the highest and 
the lowest fitness countries, we here explore the complete range of fitness 
values. This is equivalent to adding a further dimension to the analysis. The 
results are reported in figure \ref{fig:IGvsGDP_3D}. To represent the three dimensions, the 
dependent variable, the growth due to inputs, is visualized as a color map. To further explain the 
relation between the two representations, we notice that the leftmost part of 
figure \ref{fig:IGvsGDP_3D} is populated by the same countries belonging to the dashed line in 
figure \ref{fig:GDPgrowthvsGDP_Bin} (low growth potential countries), while the rightmost part is populated 
by the countries belonging to the solid line (high growth potential countries).  
This analysis strongly supports our argument: the complexity and diversification 
of a country’s economy acts as a catalyst in triggering the transition by 
significantly reducing the necessary per capita GDP. The catching up phenomenon 
is barely observable in Figure \ref{fig:GDPgrowthvsGDP} since the plot represents the average of 
different fitness levels for each level of per capita GDP, thereby mixing up 
different states of the transition. 
Even when starting from very low levels of per capita GDP, high fitness, more 
complex, countries are able to start the transition, with increasing investments 
causing increasing input growth levels. On the contrary, low fitness countries 
characterized by exports concentrated in few low complexity sectors, require 
very high levels of per capita GDP to start the transition and attract 
investments. 
It is trivial to adapt this result to a demand-side explanation: there is a 
complementarity in kindling the industrialization process of a country between 
fitness, which is a proxy for export competitiveness, and per capita GDP, which 
is a proxy for internal demand. 

%In the previous section we observed different behaviors for countries with 
%different fitness.
%A possible explanation that the fitness allows for industrialization, lowering 
%the 
%barrier to push forward the investments in inputs, has been hinted; it requires 
%however additional investigation.
%What we observed for the two categories of top and lowest fitness should be 
%generalized
%to a continuous process.
%We try this in the next exercise, in figure \ref{fig:IGvsGDP_3D}

%This plot completely supports our argument: country fitness, the complexity 
%and diversification of the economy,
%acts as a catalyst, reducing the needed GDP per capita to start the transition.
%Figure \ref{fig:GDPgrowthvsGDP} does not show a catching up behavior because 
%for each 
%level of GDP per capita the plot represents the projection
%of different fitness levels and, therefore, different states of the transition.
%
%High fitness countries are able to start the transition, the sprout of 
%investments and 
%efforts causing high input growth, even when starting from a very low level of 
%GDP per capita. 
%On the opposite, low fitness countries, characterized by exports concentrated 
%in few sectors, 
%requires very high level of GDP per capita to start the transition 
%and attract investments. 
%
%It is trivial to integrate this result in a demand side explanation: there is a 
%complementarity 
%between fitness, a proxy for
%export competitiveness, and GDP per capita, a proxy for internal demand, in 
%kindling the 
%process of industrialization of a country.

However, even a supply-side explanation is consistent with our empirical 
results, since the opening of new export sectors increases the incentives to 
invest. One would expect this increase in incentives to be even more visible 
looking at the residual productivity growth (a). However, if the new accessible 
sectors allow new - intrinsically different - inputs to be used and accumulated, 
the scale of production can increase without a corresponding increase in the 
factor productivity, consistently with an input driven growth. Low fitness 
countries with poorly diversified economies do not start the endogenous 
transition until they have reached an extremely high level of capital, staying 
stuck in the so-called poverty trap. In this sense the poverty trap can be seen 
as a 2 dimensional object in which both the population wellness (per capita GDP) 
and the complexity of the national economy (country fitness) play a role.

%Even a supply side explanation is however consistent with this empirical 
%results: if the new accessible
%sectors allows new - intrinsically different - inputs to be used and accumulated, 
%the scale of production
%can increase without an increase in the factor productivity, consistently with 
%an input lead
%growth.
%Going back to the basic growth model sketched in section \ref{sec:Growth}, the 
%empirical
%observation is consistent with a model connecting the fitness of the country to 
%the position 
%of $K^*_3$ in figure \ref{fig:multipleequilibria} or, equivalently, to the 
%value of 
%$K_F$ in equation \ref{eq:QuantCDGrowth}: 
%in this hypothetical model 
%high fitness countries have high $K_F$, representing the fact that in countries
%with many opportunities the population starts to invest just after the 
%subsistence
%level, and therefore $K^*_3$ is very near
%to $K^*_1$; low fitness countries, on the contrary, have low $K_F$, representing
%the fact that in countries with few opportunities the population does not 
%invest 
%until after achieving higher levels of luxury, so that $K^*_3$ is near to 
%$K^*_2$.
%Therefore they do not start 
%the endogenous transition until they have already high level of capital.

A different point of view to explain the same empirical data would be a demand
side explanation imagining a continuous transition between internal and 
external demand,
proxied respectively by the GDP per capita and the fitness of the country, that 
is a
measure of the complexity and value added of the exported goods.\\
Finally we add some comments about the trajectories of the countries we considered in Fig.\ref{fig:IGvsGDP_3D}. The depicted countries are an the extended set of Asian Tigers, originally including South Korea, Singapore, Hong Kong and Taiwan, a well known set of countries studied for their model of industrialization. In particular, South Korea and Thailand represent a paradigm for what we intend with a \textit{lateral escape} from poverty trap. Taking advantage of the two-dimensional structure of the trap, they have increased their fitness first, and when their economy reached a high level of complexity they started the upward movement towards a balanced and richer society, reaching a higher level of GDP per capita. The careful reader will note that the case of China is complementary: even if the fitness of China was high even in the sixties, the extreme poverty of a large share of its population prevented this country to overcome the barrier and start to grow in a sustained way. From this point of view, the industrialization of China from the sixties to nowadays can be described in 
terms of the neoclassical poverty trap, in which only one dimension is needed. However, this consideration can only move our story back to the sixties: we still need to understand how China reached, fifty years ago, such a level of Fitness. This investigation will be the subject of a future work. We end this section with an analogy between this subtle, country-dependent interplay between fitness and GDP per capita and phase transitions, such as the transition of water from the liquid state to the gaseous one. When water is heated, in general, its temperature increases. This is analogous to a country which is already out of the poverty trap, and whose efforts are entirely reflected in an increase of its GDP per capita. However, during a phase transition the absorbed heat is used to break the bonds among the molecules, and even if the temperature does not increase, this process is fundamental to change phase. Similarly, when a country exits from the poverty trap by increasing its fitness the consequences of 
its efforts do not give immediate results in terms of GDP per capita, but they are firmly grounding the requirements for a future, sustained growth.

\section{Conclusions}

We shown in the paper that simple toy models of countries' growth (in 
particular models 
assuming that all countries are 
homogeneous objects characterized only by one state variable, being it the per 
capita GDP or per capita 
physical capital) are not able to catch the different patterns of 
industrialization and, therefore,
to predict the starting point of industrialization, the moment in which they 
will come out of the poverty
trap. We also shown that the measure of Fitness is able to properly disentangle 
these different patterns,
highlighting countries that are ready to take off and to industrialize and 
countries that, even with similar
standard of living, are far from the threshold.

These findings suggest a possible role for the complexity of the economy to 
drive opportunities and attract internal or external sources of investment. The 
increased number of complex production sectors, proxied by the fitness, leads 
to an open array of possibilities allowing the individual to invest in physical 
and human capital in order to exploit new and additional opportunities. This 
effect of fitness on savings and education could be modeled in terms of a 
dynamical process in which a multi-sector economy goes from one equilibrium to 
another after overcoming a threshold that can be lowered by increasing the 
complexity and diversification of the economy. We believe that this out of 
equilibrium dynamics is the one performed by countries emerging from the 
poverty trap. 

Indeed, this analysis does not scrap the concept of poverty trap: if anything, 
it makes the original point stronger. As can be seen from figure \ref{fig:IGvsGDP_3D}, not only is 
the barrier real but, for countries with low fitness, it is extremely high; to 
the point that the figure seems to imply the paradox that in order emerge from 
the poverty trap a low fitness country must first become rich. However, it also 
suggest a different way to overcome both the trap and the paradox, lowering the 
threshold to industrialization by diversifying exports and making its economy 
more complex. 

We think that our analysis is particularly relevant for policy makers 
interested in the first steps of development, in particular for the ones 
interested in countries that are unable to complete their industrialization 
process even if they enjoy moderately good standards of living thanks to the 
presence of natural resources.

\section{Acknowledgments}

The authors acknowledge funding from the “CNR Progetto di Interesse CRISIS LAB” 
(http://www.crisislab.it) and EU Project no. 611272 GROWTHCOM 
(http://www.growthcom.eu) The funders had no role in study design, data 
collection and analysis, decision to publish, or preparation of the manuscript. 
We thank Matthieu Cristelli and Andrea Tacchella for useful discussions and 
Fabio Saracco for discussions and data processing. 

\bibliographystyle{plainnat}
\bibliography{growth,complexity}
\appendix

\section{Dataset\label{sec:dataset}}

In the following we will use mostly simple national statistics, other than the 
Fitness measure previously described.

We use for all the national statistics data the Penn World Table 8.0.
The data on physical capital is produced with a perpetual inventory method, as 
described in \cite{inklaar2013capital},
while the proxy for the human capital is the average number of years of 
education in the population.
The exogenous growth has been computed as the residual of the total growth 
after 
removing the input growth.

Countries' fitness evaluations are based on the import-export flows as 
registered in the UN-NBER database, reconstructed and edited by 
\cite{finestra}. This database includes the (discounted) imports of 72 
countries 
and covers more than 2577 product categories for a period ranging from 1963 to 
2000. Exports are reconstructed starting from these imports, which cover about 
the 98\% of the total trade flow. There are many possible categorizations for 
products; we will base our study on the Sitc v2, 4-digits coding. After a data 
cleaning procedure, whose aim is to remove obvious errors in the database 
records, and to obtain a consistent collection of data, the number of countries 
fluctuates between 135 and 151 over the years, while the number of products 
remains equal to 538.

\section{Basic Solow Model of Growth and Growth Accounting}\label{apx:solow}

In this appendix we give a simple version of the models of economic growth based 
on \cite{solow1956contribution} and \cite{solow1957technical}. 
Even if the aim of the paper is an 
empirical analysis, we thought it might be useful to the potential reader 
new to the macroeconomic analysis to have a simple version of economic model 
aimed at explaining growth.
Moreover this derivation is required in our analysis to decompose the growth 
in input growth and exogenous growth.

We start writing a production function as generic as possible,
\begin{equation}
Y_{i,t}=Y(A_{i,t},I_{i,t}^j),
\end{equation}
where $Y_{i,t}$ is the production of country $i$ at time $t$, $A_{i,t}$ is 
an efficiency measure and, for different $j$s, $I_{i,t}^j$ are the different 
inputs of the production 
(Physical Capital, Labor, Human Capital, ...). The production function $F$ 
gives the output of the economy for different levels of inputs and 
efficiency.
The growth of output, that we will identify with GDP in the following, can 
therefore be the consequence of an efficiency and technological growth, i.e. a
growth of $A$, or an input growth. 

In growth models some inputs are accumulated in an endogenous way: a part of 
the output is invested to build new physical capital, a part of the working 
time 
of the laborers is spent to train new workers and accumulate human capital.
Their level in equilibrium is the result of their accumulation and their 
depreciation. Growth due to input accumulation is therefore called endogenous 
growth.
On the opposite side, growth due to technology and efficiency, a growth of 
$A$, is 
called exogenous growth\footnote{it is also called exogenous growth the 
growth of inputs that are not accumulated endogenously in the model; e.g. 
demographic growth}.

In the following we will use a minimal case. We will take a Cobb-Douglas 
production 
function with two inputs, physical capital $K_{i,t}$ and labor $L_{i,t}$:
\begin{equation}
Y_{i,t}=A_{i,t}K_{i,t}^{\alpha}L_{i,t}^{1-\alpha}.\label{eq:CDGrowth}
\end{equation}

\begin{figure}[!ht]
\begin{subfigure}[]{
\includegraphics[width=8cm]{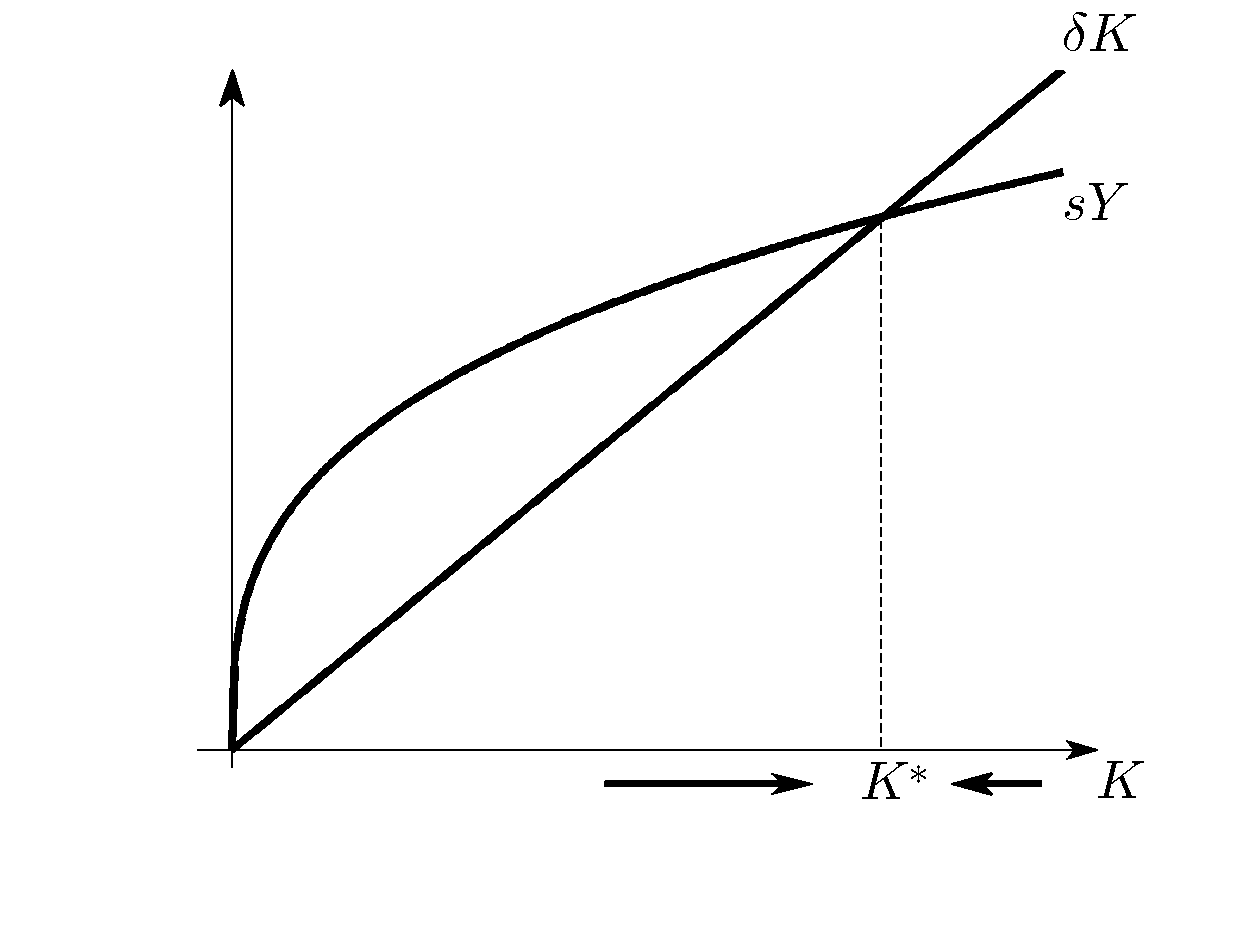}}
\end{subfigure} 
\quad
\begin{subfigure}[]{
\includegraphics[width=8cm]{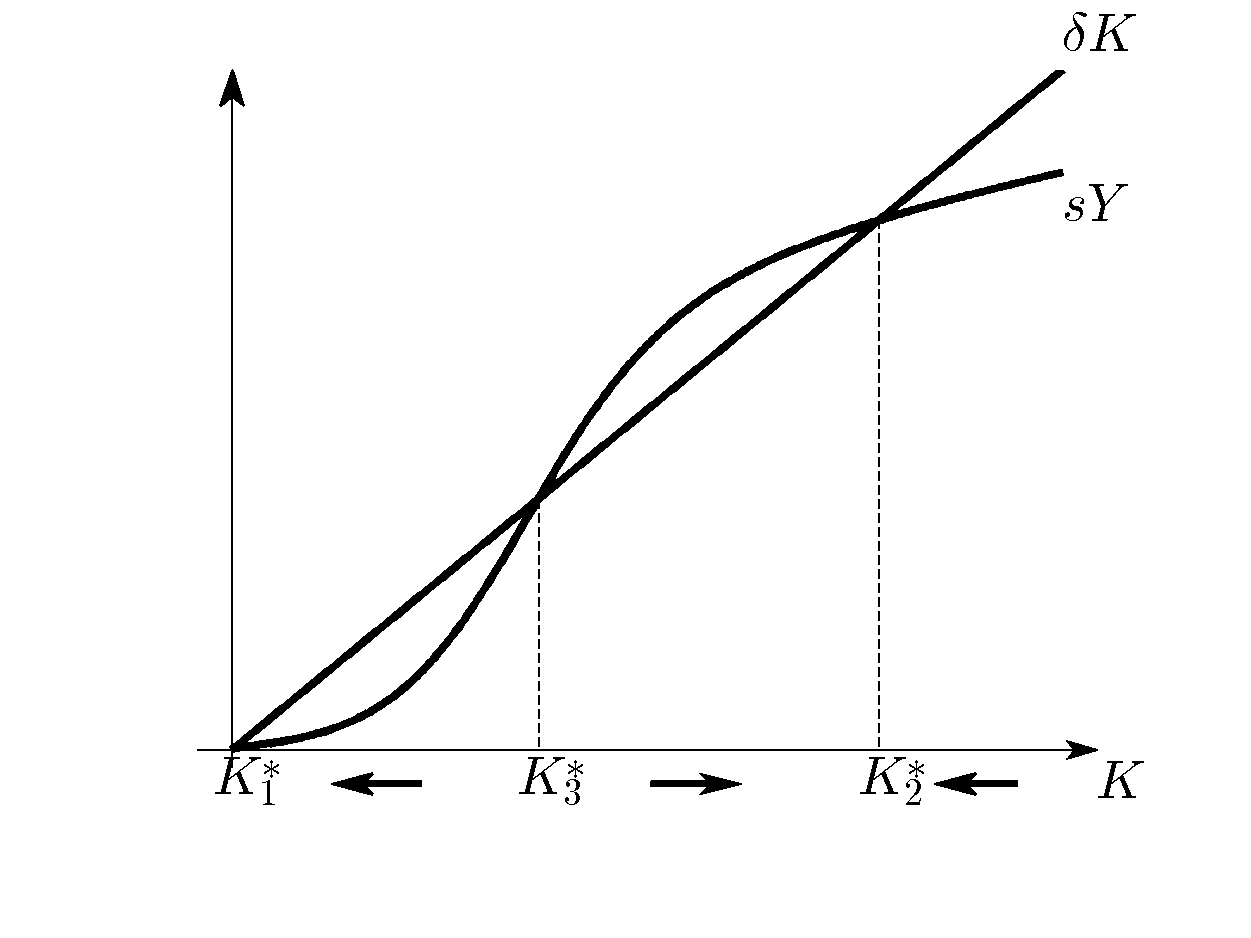}}
\caption{(a): a single equilibrium in $K^*$. The two curves $\delta K$ and $sY$ 
represents respectively capital
depreciation and capital accumulation. The equilibrium is achieved when the 
two effects are equal, i.e. $K^*$. (b): a case with multiple equilibria. When the production function has 
increasing returns for some scales,
multiple equilibria are possible. 
$K^*_1$ and $K^*_2$ are stable equilibria, while $K^*_3$ is unstable.} 
\label{fig:multipleequilibria}
\end{subfigure}
\end{figure}

If a fraction $s$ of the output $Y$ is invested in the production of new 
physical 
capital $K$ and a fraction $\delta$ of $K$ decades at each time step due to 
depreciation, the time evolution of physical capital is
\begin{equation}
K_{i,t+1}=sY_{i,t}+(1-\delta)K_{i,t}.
\end{equation}
In figure \ref{fig:multipleequilibria}(a) we show that this equation has only one 
stable equilibrium.
The equilibrium is possible due to the decreasing returns on capital: the more 
capital a country has, the less
output the country gains with an additional unit of capital.
It is also worth to note that, since the equilibrium point $K^*$ depends on 
$A$ and $L$, $K$ will still grow if those variables grow.

However already in \cite{solow1956contribution} there were the 
idea 
that multiple equilibria
in the capital levels are possible. 
If for example the investing rate $s$ is not independent from the per capita 
income
of the country, but it depends on the achievement on a minimal level
of subsistence, the capital accumulation function becomes non linear.
We can for example assume a functional form of the investing rate like,
\begin{equation}
s_{i,t}=\frac{s_i}{1+\mathrm{e}^{K_F-K_{i,t}}}
\label{eq:QuantCDGrowth}
\end{equation}
where $s_i$ is a country-dependent parameter and $K_F$ a minimum threshold to 
achieve 
subsistence and start investing.
As a consequence of this non linearity the system can 
present a behavior similar to figure \ref{fig:multipleequilibria}(b).

After overcoming a barrier, in figure \ref{fig:multipleequilibria}(b) represented 
by point
$K^*_3$, the country capital would move endogenously to $K^*_2$.
The out-of-equilibrium dynamics from one 
equilibrium to another has to be characterized by fast input accumulation and,
therefore, fast endogenous economic growth.
A way of looking the results of our analysis in section \ref{sec:model}
is that $k_F$ depends on the Fitness of the country.

In this setting we can also understand equation \ref{eq:GrowthDec}. From \ref{eq:CDGrowth},
the GDP per capita is equal to 
\begin{equation}
\left(\frac{Y_{i,t}}{P_{i,t}}\right)_{i,t}=A_{i,t}\left(\frac{K_{i,t}}{P_{i,t}}
\right)^\alpha\left(\frac{E_{i,t}}{P_{i,t}}H_{i,t}\right)^{1-\alpha},
\end{equation}
and therefore, defining with the lowercase letters the growth rates of the 
respective 
uppercase variables and with the hat the division by population,
\begin{equation}
 \hat{y}_{c,t}=a_{c,t}+\alpha\hat{k}_{c,t} + (1-\alpha)\hat{e}_{c,t} + 
(1-\alpha) h_{c,t}.
\end{equation}
In equation \ref{eq:GrowthDec} we dropped the hats notation to simplify the reading, and 
$y$, $k$, and $e$ are presented directly as the GDP per capita, physical capita per capita,
and employment rate.

\end{document}

%% file: abstract.tex
% The analysis of inputs growth rates along the industrialisation patterns of 
% countries shows that economic complexity plays a major role in characterising % 
% the dynamics. In particular high fitness, more differentiated, and complex 
% economies, face a lower barrier to start the transition through 
% industrialisation.
We analyze the decisive role played by the complexity of economic systems at the onset of the industrialization process of countries over the past 50 years. Our analysis of the input growth dynamics, based on a recently introduced measure of economic complexity, reveals that more differentiated and more complex economies face a lower barrier (in terms of GDP per capita) when starting the transition towards industrialization. Moreover,
adding the complexity dimension  to the industrialization process description helps to reconcile current theories with empirical findings.